\documentclass[twocolumn,fleqn,english,aps,prd,preprintnumbers,
showkeys,nofootinbib,superscriptaddress,floatfix,tightenlines]{revtex4}  
\usepackage[T1]{fontenc}
\usepackage[utf8]{inputenc}
\usepackage{amsmath}
\usepackage{amsthm}
\usepackage{amssymb}
\usepackage{graphicx}
\usepackage{amsfonts}
\usepackage{amscd}
\usepackage{amsxtra}
\usepackage{epstopdf}
\usepackage{color}
\usepackage{dcolumn}
\usepackage{bm}
\usepackage{multirow}
\usepackage{slashed}

\makeatletter

\usepackage[normalem]{ulem}
\usepackage[dvipsnames]{xcolor}
\usepackage{array}
\usepackage{slashed}
\renewcommand{\sout}{\bgroup \color{red} \ULdepth=-.5ex \ULset}

\makeatother

\normalsize

\usepackage{babel}
\begin{document}
\preprint{INHA-NTG-12/2022}
\title{Medium modification of singly heavy baryons in a pion-mean
  field approach} 
\author{Nam-Yong Ghim}
\email{namyong.gim@inha.edu}
\affiliation{Department of Physics, Inha University, Incheon 22212,
Republic of Korea} 

\author{Hyun-Chul Kim}
\email{hchkim@inha.ac.kr}
\affiliation{Department of Physics, Inha University, Incheon 22212,
  Republic of Korea} 
\affiliation{School of Physics, Korea Institute for Advanced Study 
  (KIAS), Seoul 02455,  Korea}

\author{Ulugbek Yakhshiev}
\email{yakhshiev@inha.ac.kr}
\affiliation{Department of Physics, Inha University, Incheon 22212,
 Republic of Korea}
\affiliation{Theoretical Physics Department, National University
of Uzbekistan, Tashkent 100174,
  Uzbekistan}

\author{Ghil-Seok Yang}
\email{ghsyang@hoseo.edu}
\affiliation{Department of General Education for Human Creativity,
  Hoseo University, Asan 31499, Republic of Korea} 

\selectlanguage{english}%
\begin{abstract}
We investigate how the masses of singly heavy baryons undergo changes
in nuclear matter. The mass spectrum of the singly heavy baryons was
successfully described in a pion-mean field approach even with isospin 
symmetry breaking, based on which we extend the investigation to the
medium modification of the singly heavy baryons. Since all dynamical
parameters were determined by explaining the mass spectrum of the
SU(3) light and singly heavy baryons in free space, we can directly
implement the density-dependent functionals for the dynamical
parameters, of which the density dependence was already fixed by
reproducing the bulk properties of nuclear matter and medium
modification of the SU(3) light baryons. We predict and discuss the
density dependence of the masses of the singly heavy baryons. 
\end{abstract}
\maketitle

\section{Introduction}
Understanding hadrons in nuclear medium have been one of the most
important issues in hadronic and nuclear physics, since it is deeply
connected to nonperturbative aspects of quantum chromodynamics (QCD):
the restoration of chiral symmetry and quark
confinement~\cite{Drukarev:1991fs, Birse:1994cz, Brown:1995qt,
  Saito:2005rv}. The quark condensate, an order parameter of
spontaneous breakdown of chiral symmetry, is known to decrease in
nuclear medium, which indicates that chiral symmetry tends to be
restored as the nuclear density
increases~\cite{Drukarev:1991fs}. Experimentally, it has also been 
observed that the mass of the nucleon undergoes change in
nuclei~\cite{EuropeanMuon:1983wih, JeffersonLabE93-049:2002asn,
  HADES:2014lrq, Malace:2014uea, Eskola:2016oht, A1:2020clw}. It
implies that other baryons may be modified in nuclear matter. 
In the present work, we want to focus on how the masses of the singly 
heavy baryons changes in nuclear matter. The heavy flavors in nuclei
were already investigated right after the $J/\psi$ was
found~\cite{Tyapkin:1975, Tyapkin:1976, Iwao:1976yi, Dover:1977hs,
  Dover:1977jw, Bando:1981ti}. The singly heavy baryons $\Lambda_c$ and
$\Sigma_c$ in nuclear matter was examined in relativistic mean-field
theory~\cite{Tan:2004mt}, the quark-meson coupling
model~\cite{Tsushima:2002cc, Tsushima:2002ua, Tsushima:2003dd}, and
QCD sum rules~\cite{Wang:2011hta, Wang:2011yj, Wang:2012xk,
  Azizi:2016dmr}  (see also a recent review and references
therein~\cite{Hosaka:2016ypm}). Recently, the SU(3) Skyrme model with
a bound-state approach was applied to the masses of the singly heavy
baryons in nuclear matter~\cite{Won:2021pwb}. 

Recently, we have investigated how the masses of the SU(3) baryons
undergo changes in nuclear medium, based on the medium-modified pion
mean-field approach~\cite{Ghim:2021odo}. We have first examined
baryonic matter including symmetric matter, asymmetric matter, neutron
matter, and strange baryonic matter, taking empirical information on
the bulk properties of nuclear matter as a guiding principle. By
describing the various matters and masses of the SU(3) baryons in
nuclear medium, we were able to fix all density-dependent
parameters. Thus, we can proceed to study the masses of the singly
heavy baryons in nuclear medium with parameters already fixed. The
pion mean-field approach, also known as the chiral quark-soliton model
($\chi$QSM), was constructed by Witten's seminal
idea~\cite{Witten:1979kh}: in the large $N_c$ (the number of colors)
limit, the nucleon can be regarded as a state of $N_c$ valence quarks
bound by the pion mean field generated self-consistently by the
presence of the $N_c$ valence quarks. The same idea can be applied to
the singly heavy baryons. If we take the limit of the infinitely
heavy-quark mass ($m_Q\to\infty$), a heavy quark resided in a singly
heavy baryon can be decoupled from the $N_c-1$ valence quarks inside
it. Thus, the heavy quark inside a singly heavy baryon is considered
as a mere static color source and the quark dynamics inside it is
governed by the light quarks. Since the heavy quark is infinitely
heavy, the heavy-quark spin is conserved, which leads to the
conservation of the light-quark spin. It is known as the heavy-quark
spin symmetry. In this heavy-quark mass limit, the singly heavy baryon
is independent of the heavy flavor, which is called the heavy-quark
flavor symmetry~\cite{Isgur:1989vq, Isgur:1991wq, Georgi:1990um}. In
this picture, the singly heavy baryons are 
represented by a baryon antitriplet ($\overline{\bm{3}}$) and two
baryon sextets ($\bm{6}$) with spin 1/2 and 3/2. Thus, the singly
heavy baryons can be considered as a bound state of the $N_c-1$
valence quarks with the single heavy quark detached. The heavy quark
is required only for making the singly heavy baryon a color singlet. 

Based on this idea, the pion mean-field approach was directly extended
to the singly heavy baryons~\cite{Yang:2016qdz}. It has successfully
described various properties of the singly heavy baryons in free
space~\cite{Kim:2017khv, Kim:2018xlc, Kim:2018nqf, Yang:2018uoj,
  Yang:2019tst, Kim:2019wbg, Yang:2020klp, Kim:2020uqo, Kim:2020nug,
  Kim:2021xpp, Suh:2022atr, Suh:2022ean} (see also a
recent review~\cite{Kim:2018cxv}). As mentioned previously, using the 
pion mean-field approach, we were able to describe how the masses of
the SU(3) baryons are modified in nuclear
medium~\cite{Ghim:2021odo}. The bulk properties of nuclear matter
evaluated from the present approach were in good agreement with
empirical and experimental data. We proceed now to describing the
masses of the singly heavy baryons with both spin 1/2 and 3/2. 

The paper is organized as follows: In the next Section, we
briefly review the general formalism for the pion mean-field
approach. In Section~\ref{sec:3}, we show how to implement the
density-dependence into the dynamical parameters.  
In Section~\ref{sec:4} we present the numerical results and discuss them.
The last Section is devoted to the summary and conclusions of the
present work. The explicit expressions for the baryon masses are
presented in Appendix\,\ref{app:a}. 

\section{General formalism}
\label{sec:2}

The pion mean-field approach allows one to describe both the light and
singly heavy baryons on an equal footing. Replacing one light quark by
a heavy quark with the infinitely heavy mass, we can construct a state
for the singly heavy baryon~\cite{Kim:2021xpp}. We first define the
normalization of the baryon state in the large $N_c$ limit as 
\begin{align}
\langle B(p', J_3')| B(p,J_3)\rangle = 2 M_B \delta_{J_3' J_3}
  (2\pi)^3 \delta^{(3)} (\bm{p}'-\bm{p}),   
\end{align}
where $M_B$ denotes the corresponding baryon mass. A state of the
singly heavy baryon is then expressed as 
\begin{align}
|B,p\rangle &= \lim_{x_4\to-\infty} \exp(ip_{4}x_{4})
              \mathcal{N}(\bm{p}) \cr 
&\hspace{-1.cm} \times \int d^3 x  \exp(i\bm{p}\cdot
  \bm{x})(-i\Psi_h^\dagger(\bm{x}, 
  x_4) \gamma_4) J_B^\dagger (\bm{x},x_4) 
              |0\rangle,\cr
\langle B,p| &= \lim_{y_4\to \infty} \exp(-ip'_4 y_4) 
               \mathcal{N}^*(\bm{p}') \cr
& \hspace{-1.cm} \times \int d^3 y
              \exp(-i\bm{p}'\cdot \bm{y}) 
  \langle 0| J_{B}
                 (\bm{y},y_4) \Psi_h (\bm{y},y_4), 
\end{align}
where $\mathcal{N}(\bm{p}) (\mathcal{N}^*(\bm{p}'))$ denotes
the normalization factor depending on the initial (final) momentum. 
$J_B(x)$ and $J_B^\dagger(y)$ represent the Ioffe-type current of the 
$N_c-1$ valence quarks~\cite{Diakonov:1987ty} defined by 
\begin{align}
J_B(x) &= \frac1{(N_c-1)!} \epsilon_{\alpha_1\cdots \alpha_{N_c-1}} 
\Gamma_{(TT_3Y)(JJ_3Y_R)}^{f_1\cdots  f_{N_c-1}} \cr
&\hspace{-1.cm} \times \psi_{f_1 \alpha_1}(x)\cdots \psi_{f_{N_c-1}
  \alpha_{N_c-1}}(x),\cr   
J_{B}^\dagger(y) &= \frac1{(N_c-1)!} \epsilon_{\alpha_1\cdots \alpha_{N_c-1}}
\Gamma_{(TT_3Y)(JJ_3'Y_R)}^{f_1\cdots f_{N_c-1}} \cr
&\hspace{-1.cm} \times (-i\psi^\dagger(y)\gamma_4)_{f_1\alpha_1} \cdots
  (-i\psi^\dagger(y)\gamma_4)_{f_{N_c-1}\alpha_{N_c-1}} ,
\end{align}
where $f_1\cdots f_{N_c-1}$ and $\alpha_1\cdots\alpha_{N_c-1}$
designate respectively the spin-isospin and color
indices. The matrices $\Gamma_{(TT_3Y)(JJ_3Y_R)}$ carry the quantum
numbers $(TT_3Y)(JJ_3Y_{R})$ for the corresponding baryon. 
$\psi_{f_k \alpha_k}(x)$ denotes the
light-quark field and $\Psi_h(x)$ stands for the heavy-quark field. 
In the limit of $m_Q\to \infty$, a singly heavy baryon satisfies the
heavy-quark flavor symmetry. Then the heavy-quark field can be written
as 
\begin{align}
\Psi_h(x) = \exp(-im_Q v\cdot  x) \tilde{\Psi}_h(x), 
\end{align}
where $\tilde{\Psi}_h(x)$ is a rescaled heavy-quark field almost on
mass-shell. It carries no information on the heavy-quark mass in the
leading order approximation in the heavy-quark expansion. $v$ denotes
the velocity of the heavy quark~\cite{Isgur:1989vq, Isgur:1991wq,
  Georgi:1990um}. 

We now prove that the normalization factor
$\mathcal{N}^*(\bm{p}')\mathcal{N}(\bm{p})$ is correctly reduced to  
$2 M_B$, which can be computed as 
\begin{align}
 \label{eq:normal}
\langle B(p',J_3') | B (p,J_3)\rangle &=
\frac{1}{\mathcal{Z}_{\mathrm{eff}}}
  \mathcal{N}^*(p')\mathcal{N}(p) \cr
& \hspace{-2.cm} \times
\lim_{x_4\to -\infty}  \lim_{y_4\to \infty}   
\exp\left(-iy_4p_4'+ix_4p_4\right)  \cr
& \hspace{-2.cm}\times \int  d^3x d^3y 
  \exp(-i\bm{p}'\cdot \bm{y}+  i\bm{p}\cdot \bm{x})\cr
& \hspace{-2.cm}\times
 \int \mathcal{D} U  \mathcal{D} \psi
  \mathcal{D}\psi^\dagger \mathcal{D} \tilde{\Psi}_h \mathcal{D}
  \tilde{\Psi}_h^\dagger J_B (y) \cr
& \hspace{-2.cm}\times \Psi_h(y) (-i\Psi_h^\dagger(x) \gamma_4)
  J_B^\dagger(x)  \exp\left[\int d^4z\left\{\right.\right. \cr
&\hspace{-2.cm}\times  
  (\psi^\dagger(z))_{\alpha}^{f} \left( i\rlap{/}{\partial}  + i
  MU^{\gamma_5} + i  \hat{m} \right)_{fg} \psi^{g\alpha}(z)  \cr 
& \hspace{-2.cm} + \left. \left. 
  \Psi_h^\dagger(z) v\cdot \partial 
  \Psi_h(z) \right\} \right] \cr
&  \hspace{-2.cm}= 
  \frac{1}{\mathcal{Z}_{\mathrm{eff}}}\mathcal{N}^*(p')\mathcal{N}(p)
  \lim_{x_4\to -\infty} \lim_{y_4\to \infty} \cr
& \hspace{-2.cm}\times 
 \exp\left(-iy_4p_4'+ix_4p_4\right) \cr
& \hspace{-2.cm} \times \int  d^3x d^3y 
  \exp(-i\bm{p}'\cdot y+  i\bm{p}\cdot \bm{x}) \cr
& \hspace{-2.cm} \times \langle 
J_B (y) \Psi_h(y) (-i\Psi_h^\dagger(x) \gamma_4) J_B^\dagger(x)\rangle_0.  
\end{align}
Here, $\mathcal{Z}_{\mathrm{eff}}$ is the low-energy effective
QCD partition function defined as 
\begin{align}
\mathcal{Z}_{\mathrm{eff}} = \int   \mathcal{D} U \exp(-S_{\mathrm{eff}}).    
\label{eq:Zeff}
\end{align}
$S_{\mathrm{eff}}$ is called the effective chiral action expressed as  
\begin{align}
S_{\mathrm{eff}} = -N_c \mathrm{Tr} \ln \left[ i\rlap{/}{\partial}  +
  iMU^{\gamma_5} + i   \hat{m}\right]. 
\label{eq:seff}
\end{align} 
$\langle ... \rangle_{0}$ in Eq.~\eqref{eq:normal} expresses the
vacuum expectation value of the baryon correlation function. 
$M$ denotes the dynamical quark mass and the $U^{\gamma_5}$ represents
the chiral field defined by 
\begin{align}
  \label{eq:1}
U^{\gamma_5} (z) = \frac{1-\gamma_5}{2}   U(z) + U^\dagger(z)
  \frac{1+\gamma_5}{2} 
\end{align}
with
\begin{align}
  \label{eq:2}
U(z) = \exp[{i\pi^a(z) \lambda^a}].  
\end{align} 
$\pi^a(z)$ are the pseudo-Nambu-Goldstone (pNG) fields 
and $\lambda^a$ the flavor Gall-Mann matrices. 
$\hat{m}$ is the mass
matrix of current quarks $\hat{m} = \mathrm{diag}(m_{\mathrm{u}},\,
m_{\mathrm{d}},\,m_{\mathrm{s}})$. We treat the strange current quark
mass $m_{\mathrm{s}}$ perturbatively. The propagators of a light
quark in the $\chi$QSM~\cite{Diakonov:1987ty} is obtained to be  
\begin{align}
G(y,x) &= \left \langle y \left |
                      \frac1{i\rlap{/}{\partial} + i 
  MU^{\gamma_5}  +i\overline{m}} (i\gamma_4)\right | x\right 
\rangle \cr
&\hspace{-1cm}= \Theta(y_4   -x_4) 
\sum_{E_n>0} e^{-E_n(y_4-x_4)} \psi_n(\bm{y})
  \psi_n^\dagger(\bm{x}) \cr
&\hspace{-1cm}  - \Theta(x_4  -y_4) 
\sum_{E_n<0} e^{-E_n(y_4-x_4)}
  \psi_n(\bm{y})\psi_n^{\dagger}(\bm{x}), 
\end{align}
where $\Theta(y_4   -x_4)$ is the Heaviside step
function. We introduce $\overline{m}$, which is the average mass of
the up and down current quarks: $\overline{m}=(m_{\mathrm{u}}+
m_{\mathrm{d}})/2$. It properly generates the Yukawa tail
of the pion mean field, when we later solve the equation of
motion. We define the one-body Dirac Hamiltonian as 
\begin{align}
  H = \gamma_4 \gamma_i \partial_i + \gamma_4 MU^{\gamma^5}
    + \gamma_4 \bar{m} \mathbf{1}.
\end{align}
Solving the eigenvalue problem of $H$, we find the energy eigenvalues
corresponding to the single-quark eigenstate
\begin{align}
H\psi_n(\bm{x}) = E_n \psi_n(\bm{x}).  
\end{align}

We now deal with the heavy-quark propagator in the limit of
$m_Q\to\infty$ 
\begin{align}
G_h (y,x) = \left \langle y \left |   \frac1{\partial_4 }\right |
  x\right \rangle= \Theta(y_4-x_4) \delta^{(3)}(\bm{y} - \bm{x}) .  
\end{align}
Using these quark propagators and taking the limit of
$y_4-x_4=T\to\infty$, we evaluate the baryon correlation 
function $\langle J_B (y)\Psi_h(y) (-i\Psi^{\dagger}_h(x)\gamma_4)
J_B^\dagger(x)\rangle_0$ as follows~\cite{Diakonov:1987ty,
  Christov:1995vm}:   
\begin{align}
&\langle J_B (y)\Psi_h(y) (-i\Psi_h^\dagger(x) \gamma_4)
J_B^\dagger(x)\rangle_0 \cr
&  \sim \exp\left[-\{(N_c-1)E_{\mathrm{val}} +
  E_{\mathrm{sea}}+m_Q\}T\right] \cr
&= \exp[-M_B T],  
\label{eq:baryon_corr}
\end{align}
which cancels the term $\exp\left(-iy_4p_4'+ix_4p_4\right)= \exp[M_BT]
$ in the large $N_c$ limit. Therefore, we prove that the normalization
factor becomes $\mathcal{N}^*(\bm{p}')\mathcal{N}(\bm{p})=2M_{B}$.  
Utilizing this normalization and Eq.~\eqref{eq:baryon_corr}, we derive 
the classical mass of the singly heavy baryon~\cite{Kim:2018xlc} as    
\begin{align}
M_B = (N_c-1) E_{\mathrm{val}} + E_{\mathrm{sea}} + m_Q.  
\end{align}
Before we proceed to compute the mass spectrum of the singly heavy
baryons, we want to mention the ordering of the two limits: $N_c\to
\infty$ and $m_Q\to \infty$. We first take the limit of $m_Q\to
\infty$ and then we carry out $N_c\to \infty$. This ordering is
compatible with the present pion mean-field approach. If we had taken
the ordering inversely, we would not have detached the heavy quark
from the singly heavy baryons.  

We restate $S_{\mathrm{eff}}$ in Eq.~\eqref{eq:seff} in the
following form 
\begin{align}
S_{\mathrm{eff}}(U) \;=\; -N_{c}\mathrm{Tr}\ln iD(U)\,,
\label{eq:echl}
\end{align}
where the trace operator $\mathrm{Tr}$ runs over spacetime and all
relevant internal spaces. The $N_c$ stands for the number 
of colors, and $D(U)$ the one-body Dirac differential operator is
defined by  
\begin{align}
D(U) = \gamma_{4}(i\rlap{/}{\partial} - \hat{m} - MU^{\gamma_{5}}),
\label{eq:Dirac}  
\end{align}
where $\partial_4$ is the time derivative in Euclidean space.
The mass matrix of the current quarks $\hat{m}$ can be expressed in
terms of the Gell-Mann matrices
\begin{align}
\hat{m} = m_1 \bm{1} + m_3 \lambda_3 + m_8 \lambda_8,   
\end{align}
where 
\begin{align}
m_0 & = \frac{m_{\mathrm{u}} + m_{\mathrm{d}} + m_{\mathrm{s}}}{3}, \cr
m_3 & = \frac{m_{\mathrm{u}} - m_{\mathrm{d}}}{2},\cr
m_8 & = \frac{m_{\mathrm{u}} + m_{\mathrm{d}} -2
      m_{\mathrm{s}}}{2\sqrt{3}}. 
\end{align}
$U^{\gamma_5}$ denotes the SU(3) chiral field 
\begin{align}
  U^{\gamma_5} &= \exp[i\pi^a \lambda^a \gamma_5] =
                 \frac{1+\gamma_5}{2} U + \frac{1-\gamma_5}{2}
                 U^\dagger,
\end{align}
where $\pi^a(\bm{r})$ is the pNG field with flavor indices $a=1,\cdots
N_f^2-1$. $N_f$ is the number of flavors. Since the hedgehog symmetry
constrains the form of the classical pion field as  $\bm{\pi}(\bm{x})
= \hat{\bm{n}}\cdot \bm{\tau} P(r)$, where $P(r)$ is called the profile
function of the soliton, we keep only the pion fields $\pi^a$ with
$a=1,\,2,\,3$. Thus, we have the SU(2) chiral $U$ field as
$U_{\mathrm{SU(2)}}=\exp(i\hat{\bm{n}}\cdot \bm{\tau} P(r))$.  
We now embed the SU(2) soliton into SU(3) by Witten's
ansatz~\cite{Witten:1983tx}  
\begin{align}
U^{\gamma_{5}}(x) \;=\; \left(\begin{array}{lr}
U_{\mathrm{SU(2)}}^{\gamma_{5}}(x) & 0\\
0 & 1
\end{array}\right).
\end{align}
Since we consider the mean-field approximation, we can carry out the
integration over $U$ in Eq.~\eqref{eq:Zeff} around the saddle
point ($\delta S_{\mathrm{eff}}/\delta \pi^a =0$). This saddle-point
approximation yields the equation of motion that can be solved
self-consistently. The solution provides the self-consistent profile
function $P_c(r)$, which is just the pion mean field. Compared to the
SU(3) light baryons, it is weaker than that produced by the $N_c$
valence quarks. 

Since the classical $U_{\mathrm{cl}}$ field is not invariant under
translation and rotation, we need to restore these symmetries such
that we have the singly heavy baryons with correct quantum numbers. 
Thus, we perform the zero-mode quantization or the semiclassical
quantization for the chiral soliton. A detailed formalism for the
zero-mode quantization can be found in Ref.~\cite{Christov:1995vm,
  Diakonov:1997sj}. Having quantized the soliton, we obtain the
collective Hamiltonian as     
\begin{align}
H_{\mathrm{coll}} = H_{\mathrm{rot}} + H_{\mathrm{sb}},   
\end{align}
where the rotational part of the collective Hamiltonian 
is given as 
\begin{align}
  H_{\mathrm{rot}} 
  =  \frac{1}{2\overline{I}_{1}}\sum_{i=1}^{3}\hat{J}_{i}^{2}
  +\frac{1}{2\overline{I}_{2}}\sum_{p=4}^{7}\hat{J}_{p}^{2}.
  \label{eq:Hrot}
\end{align}
Here $\overline{I}_{1}$ and $\overline{I}_{2}$ 
have forms
\begin{align}
  \overline{I}_{1}= \eta I_{1},\quad \overline{I}_{2}= \eta I_{2}.
  \label{eq:I1I2}
\end{align}
where $I_1$ and $I_2$ are the usual moments of inertia. Since we take
an ``\emph{model-independent}'' approach~\cite{Adkins:1984cf}, we do
not compute all the dynamical parameters such as $I_1$ and $I_2$ but 
determine them by using the experimental data on the mass splitting of
the baryon octet and decuplet. In the case of the singly heavy
baryons, we only know that $\overline{I}_1$ and $\overline{I}_2$
should be smaller than $I_1$ and $I_2$ because the pion mean field
from the $N_c-1$ valence quarks is weaker than that with the $N_c$
ones.  Thus, we fit $\eta$ to the masses of the singly heavy baryons
in free space~\cite{Yang:2016qdz}. 
The $J_{i}$ are the generators of the SU(3) group of which the first
three components are the ordinary spin operators. More details can be 
found in Refs.~\cite{Yang:2010fm, Blotz:1992pw}. 

In representation  $\mathcal{R}=(p,q)$, the eigenvalues of
$H_{\mathrm{rot}}$ in Eq.\,\eqref{eq:Hrot} are given as  
\begin{align}
 E^{\mathrm{rot}}_{(p,q)} = 
&\left( \frac{1}{2\overline{I}_{1}} - \frac{1}{2\overline{I}_{2}}
  \right)J \left( J+1 \right) 
\cr
&+\frac{p^{2} + q^{2} + 3(p+q) + pq}{6\overline{I}_{2}} -
  \frac{3}{8\overline{I}_{2}} Y'^{2}, 
 \label{eq:Erot}
\end{align}
where $Y^{\prime}$ denotes the right hypercharge. 
In the case of the SU(3) light baryons, the presence of $N_c$ valence
quarks imposes a constraint on the collective Hamiltonian: $Y'=N_c/3$,
which selects allowed representations: the octet ($\bm{8}$) and
decuplet ($\bm{10}$). Since the singly heavy baryon consists of the
$N_c-1$ valence quarks, the right hypercharge is constrained to be
$Y^{\prime} = (N_{c}-1)/3$~\cite{Yang:2016qdz}  that allows the
antitriplet ($\overline{\bm{3}}$) and sextet ($\bm{6}$).
The center masses for the baryon antitriplet and sextet are then given
by
\begin{align}
  \label{eq:26}
  M_{\overline{\bm{3}}}^Q = M_{\mathrm{cl}} + \frac1{2\overline{I}_2},
  \;\;\;
  M_{\bm{6}}^Q =  M_{\overline{\bm{3}}}^Q + \frac1{\overline{I}_1}.  
\end{align}
Note that the center masses are flavor-independent. 

To describe the mass splitting in a representation, it is essential to
introduce the effects of isospin breaking and explicit flavor SU(3)
symmetry breaking. Expanding the effective chiral action to the linear
order of $\hat{m}$ and carrying out the quantization, we obtain the
symmetry-breaking part of the collective  Hamiltonian as 
\begin{align}
 H_{\mathrm{sb}}
  &=(m_{\mathrm{d}}-m_{\mathrm{u}})\cr
  &\hspace{-0.5cm} \times
  \left(\frac{\sqrt{3}}{2} \overline{\alpha} D^{(8)}_{38}(R)
    +\beta\hat{T}_{3} + \frac{\gamma}{2} \sum_{i=1}^{3} D_{3i}^{(8)}  
  \left( R \right)\hat{J}_{i} \right)\cr
  &\hspace{-0.5cm}+(m_{\mathrm{s}}-\overline{m})\cr
  &\hspace{-0.5cm} \times
    \left(\overline{\alpha} D_{88}^{(8)}(R) + \beta \hat{Y}+
    \frac{\gamma}{\sqrt{3}}
  \sum_{i=1}^{3}D_{8i}^{(8)}\left( R \right)\hat{J}_{i}\right),
  \label{eq:Hsb}
\end{align}
where the first term arises from the isospin symmetry breaking to
linear order, and the second term comes from the SU(3) symmetry
breaking also to linear order. Once we introduce the isospin symmetry
breaking, we need to include the contributions from the
electromagnetic (EM) self-energies of the soliton~\cite{Yang:2010id,
  Yang:2020klp}.  $D^{(8)}_{ij}$ denote SU(3) Wigner functions.
The parameters $\alpha$, $\beta$ and $\gamma$ are
expressed as
\begin{align}
   \overline{\alpha}&=\frac{N_{c}-1}{N_{c}} \alpha,\;
  \alpha=-\frac{2}{3}\frac{\Sigma_{\pi N}}{m_{\mathrm{u}} +
                      m_{\mathrm{d}}}-\beta,\cr 
  \beta&=-\frac{K_{2}}{I_{2}},\;
  \gamma=2\frac{K_{1}}{I_{1}}+2\beta,
  \label{eq:alphabetagamma}
\end{align}
where $K_{1}$ and $K_{2}$ designate the anomalous moments of
inertia. $\Sigma_{\pi N}$ stands for the pion-nucleon sigma term.
Note that $\alpha$ should be rescaled by $(N_c-1)/N_c$, because the
singly heavy baryon contains $N_c-1$ valence quarks, which modify the
pion mean field.  More discussion of $\overline{\alpha}$, $\beta$ and 
$\gamma$ can be found in Ref.~\cite{Yang:2016qdz}.

In the limit of $m_Q\to\infty$,  the spin 1/2 and 3/2 sextet states
are degenerate. To remove the degeneracy, we have to introduce the
hyperfine chromomagnetic interaction (spin-spin interaction) to order
$1/m_Q$ 
\begin{align}
  H_{LQ}^{\mathrm{HF}} 
 =\frac{2}{3}\frac{\kappa}{m_{Q}M_{\mathrm{cl}}} S_{\mathrm{L}}\cdot
  S_{\mathrm{Q}} 
 =\frac{2}{3} \frac{\varkappa}{m_{q}} S_{L}\cdot S_{Q}
  \label{eq:HLQ},
\end{align}
where the $\kappa$ stands for the anomalous chromomagnetic 
moment. The operator $S_{L}$ and $S_Q$ designate respectively the spin
operators for the soliton and heavy quark.
Taking into account the hyperfine mass splitting, the center mass of
the sextet in Eq.~\eqref{eq:26} can be decomposed into those for the
spin 1/2 and spin 3/2
\begin{align}
  \label{eq:30}
  M_{\bm{6}_{1/2}}^Q &=  M_{\bm{6}}^Q - \frac23
                       \frac{\varkappa}{m_Q},\cr 
  M_{\bm{6}_{3/2}}^Q &=  M_{\bm{6}}^Q +\frac13
                                            \frac{\varkappa}{m_Q}.
\end{align}
In addition to the EM self-energies of the soliton for the effects of
the isospin symmetry breaking, we introduce the EM interaction between
the soliton and the heavy quark, which can be formulated in the
following expression 
\begin{align}
  H_{LQ}^{\mathrm{Coul}} = \alpha_{LQ}
  \hat{Q}_{L} \hat{Q}_{Q},
  \label{eq:Hcoul}
\end{align}
where the $\hat{Q}_{L}$ and $\hat{Q}_{Q}$ represent charge operators
acting on the soliton and heavy quark. The parameter $\alpha_{LQ}$
includes the expectation value of the inverse distance  
between the soliton and heavy quark, and the fine structure constants.
We can fix it by reproducing the existing data on the masses of the
singly heavy baryons~\cite{Yang:2016qdz}.

Since almost all the dynamical parameters have already been fixed in
the light baryon sector, and their density dependences have also been
set up in the previous work~\cite{Ghim:2021odo}, we will proceed
directly to the masses of the singly heavy baryons in baryonic
matter.

\section{Singly heavy baryons in baryonic matter} 
\label{sec:3}
We now recapitulate the formalism with which we have described bulk
properties of various baryonic matters, and the masses of the SU(3)
light baryons~\cite{Ghim:2021odo}. we introduce three
density-dependent free parameters $\lambda$, $\delta$, and
$\delta_{s}$, which are respectively related to the normalized density
of infinite nuclear matter, the parameter for isospin asymmetry, and
that for the strangeness mixing.   
They are defined as  
\begin{align}
  \lambda=\frac{\rho}{\rho_{0}}, \quad  \delta=\frac{N-Z}{A},\quad
  \delta_{s} =\frac{N_{s}}{A} , 
  \label{eq:delta,detlas,lambda}
\end{align}
where the $\rho_{0}$ stands for the normal nuclear matter density,
$N$ is the number of neutrons, $Z$ the number of protons, $A$ the
baryon number, and $N_s$ the number of baryons with the strangeness 
$s=|S|$.  The strangeness is only an external free parameter, of which
the fraction identifies strange matter. We introduce 
the strangeness-mixing parameter $\chi$, which is defined as 
$\delta_s = s\chi$ such that we do not need to concern specific
strange particles that consist of strange matter. Thus, by taking the
nonzero value of $\chi$, we can consider the strange matter. 

Following Ref.~\cite{Ghim:2021odo}, we have the following
density-dependent classical mass, moments of inertia, effects of
isospin and SU(3) symmetry breaking: 
\begin{align}
  M_{\mathrm{cl}}^{*}=&M_{\mathrm{cl}} f_{\mathrm{cl}}(\lambda,
                        \delta, \delta_{1},\delta_{2},\delta_{3}) ,\\ 
  \overline{I}_{1}^{*}=&\overline{I}_{1} f_{1}(\lambda, \delta,
                         \delta_{1}, \delta_{2},\delta_{3}),\\
  \overline{I}_{2}^{*}=&\overline{I}_{2} f_{2}(\lambda, \delta,
                         \delta_{1}, \delta_{2},\delta_{3}),\\
  E_{\mathrm{iso}}^{*}
  =&\left( m_{d}-m_{u} \right)\frac{K_{1,2}}{I_{1,2}}
  f_{0}\left( \lambda,\delta,\delta_{1}, \delta_{2}, \delta_{3} \right),
  \label{eq:isosm}\\
  E_{\mathrm{str}}^{*}=&(m_{s}-\overline{m}) \frac{K_{1,2}}{I_{1,2}}
                         f_{s}(\lambda,
                         \delta,\delta_{1},\delta_{2},\delta_{3}), 
  \label{eq:strm}
\end{align}
where $f_{\mathrm{cl}}$, $f_{0,1,2}$, and $f_{s}$ are given as the
functions of the baryon density and other medium variables. 
They are explicitly written as 
\begin{align}
  f_{\mathrm{cl}}(\lambda)&=\left(1+C_{\mathrm{cl}}\lambda\right),\\
  f_{1,2}(\lambda)&=\left(1+C_{1,2}\lambda\right),
\label{eq:den_ftcl12}\\
  f_0(\lambda,\delta)
  &=1+\frac{C_{\mathrm{num}}\lambda\,\delta}{1+C_{\mathrm{den}}
  \lambda}\,,
  \label{eq:f0}\\
  f_{s}(\lambda,\delta_{s})&=1+g_{s}(\lambda)\delta_{s}\\
  g_s(\lambda) &=  sg(\lambda),\\
  g(\lambda)&= \left(6\frac{K_2}{I_2}+\frac{K_1}{I_1}\right)^{-1} \cr
              & \hspace{-1cm} \times 
              -\frac{5(M_{\mathrm{cl}}^{\ast}-M_{cl}+
 E_{(1,1)1/2}^*-E_{(1,1)1/2})}{3(m_s-\hat m)}
\label{eq:gs}
\end{align}
Once these functions are plugged in the equations of state,
the bulk properties of nuclear matter are well described up to the
density $\sim 3\rho_{0}$.  The parameters for the nuclear environment
are fixed as follows~\cite{Ghim:2021odo}: 
\begin{align}
  &C_{\mathrm{cl}}=-0.0561,\quad C_{1}= 0.6434, \quad C_{2} =
    -0.1218,\cr 
  &C_{\mathrm{num}}=65.60,\quad C_{\mathrm{den}}=0.60,
\end{align}
where $C_{\mathrm{cl}}$, $C_{1}$ and $C_{2}$ were determined
by using the empirical data on the volume energy, pressure at the
saturation point, and compressibility for symmetric nuclear matter. 
The volume energy is known to be $a_{V}=-16\ \mathrm{MeV}$  
from the semi-empirical  Bethe-Weizs\"{a}ker
formula~\cite{Bethe:1936zz,Weizsacker:1935bkz}. 
The stability condition for nuclear matter 
requires the pressure to vanish, i.e. $P=0$ near the saturation
point. The compressibility for nuclear matter 
were predicted to be $K_{0}\simeq (290\pm70)\
\mathrm{MeV}$ within various frameworks~\cite{Sharma:1988zza,
  Shlomo:1993zz, Ma:1997zzb, Vretenar:2003qm, TerHaar:1986xpv,
  Brockmann:1990cn}, whereas a slightly lower value of $K_0$, i.e.
$K_0\sim (240\pm 20)$ was suggested from the data on the energies of
the giant monopole resonance in even-even ${}^{112-124}$Sn and 
${}^{106,100-116}$Cd~\cite{Shlomo:2006}, and from earlier data on
$58\le A \le 208$ nuclei in Ref.\,\cite{Stone:2014wza}. 

The parameters $C_{\mathrm{num}}$ and $C_{\mathrm{den}}$ are relevant 
to the effects of isospin symmetry breaking. Since asymmetric
nuclear matter appears when isospin symmetry is broken, The
$C_{\mathrm{num}}$ and $C_{\mathrm{den}}$ play an essential role in
reproducing the properties of asymmetric nuclear matter such as the 
symmetry energy $\varepsilon_{\mathrm{sym}}(\lambda)$ and the slope
paramter $L_{\mathrm{sym}}$.  The symmetry energy at the saturation
density $\varepsilon_{\mathrm{sym}}(\lambda=1)$ is known to be in the
range of $\sim30-34\,\mathrm{MeV}$. The slope parameter taken
from the experiments of $^{68}$Ni, $^{120}$Sn, and $^{208}$Pb for the
neutron skin thickness indicates that heavier the nucleus is, 
larger the value of $L_{\mathrm{sym}}$~\cite{Roca-Maza:2015eza} is
observed. Thus, we choose the values of symmetry energy  
and slope parameter as $\varepsilon_{\mathrm{sym}}=32\,\mathrm{MeV}$
and $L_{\mathrm{sym}}=60\,\mathrm{MeV}$. For more details about the
medium functions and their parameters, we refer to
Ref.~\cite{Ghim:2021odo}.  

The medium modification of the EM part of the isospin symmetry
breaking is found to be small~\cite{Meissner:2008mr} and, therefore,
we ignore it in the present work. However, the spin-spin interaction
given in Eq.~\eqref{eq:HLQ} contains the classical mass of the
nucleon, which varies in nuclear medium. So, we consider its medium
modification and redefine the ratio of $\kappa$ and
$M_{\mathrm{cl}}^*$ as $\varkappa^{*}$
\begin{align}
\varkappa^{*}=\frac{\kappa}{M_{\mathrm{cl}}^*}.
  \label{eq:den_chi}
\end{align}
We neglect the medium dependence of $\kappa$, since it is only
involved in lifting the degeneracy in the baryon sextet. 
\section{Results and discussions}
\label{sec:4} 

Since all the parameters were already fixed in the light-baryon
sector, we can straightforwardly evaluate the masses of the baryon
antitriplet and sextet.  In Table~\ref{tab:1}, we list the results for
the masses of the singly charmed baryons. In the fourth column, we
first present their numerical values in free space, which have been
derived in Ref.~\cite{Yang:2020klp}. They are in remarkable agreement
with the experimental data.  
\begin{table*}[htp]
\caption{Masses of the singly charmed baryons in free space and in  
  different baryonic matters at the normal nuclear matter density
  $\lambda=1$.  The experimental data are taken from the
  PDG~\cite{ParticleDataGroup:2022pth}.
  In the fifth column, the results in symmetric nuclear matter
  ($\lambda=1$) are listed, whereas, in the sixth and seventh columns,
  those in asymmetric matter ($\delta =1$) and strange matter
  ($\chi=0.15$) are respectively given. All the masses are given in
  unit of MeV.}      
    \label{tab:1}
  \begin{center}
  \begin{tabular}{c|ccc|ccc}
  \hline \hline
Multiplet \& spin  & Baryon  & Exp. & Free space
    &\multicolumn{3}{c}{Baryonic matter  
 at $\lambda=1$.} \\
  &&&&$\delta=0,\,\chi=0$ & $\delta=1,\, \chi=0$ & $\delta=0,\, \chi=0.15$ \\
     \hline 
     \rule{0in}{2.5ex}
    & $ \Lambda_{c} $  & $2286.46\pm 0.14$ & 
    $2272.84$ & $2268.71$ & $2268.71$ & $2264.49$ \\
    $\overline{3}_{1/2}$  & $ \Xi_{c}^{+} $  & $2467.71 \pm 0.23$ & $2475.20$ & $ 2472.97$
    & $ 2411.19$ & $ 2475.09$ \\
    & $\Xi_{c}^{0} $ &  $2470.44 \pm 0.28$ & 
    $2478.18$ & $2472.16$ & $2533.94$ & $2474.27 $ \\
   \hline
   \rule{0in}{2.5ex}
   & $ \Sigma_{c}^{++} $ & $2453.91 \pm 0.14$ & $2445.67$ & $2372.37$
   & $ 2285.03$ &$2368.51 $ \\
   & $ \Sigma_{c}^{+} $ & $2452.9 \pm 0.4$ & $2444.65$ & $2370.47$
   & $2370.47$ & $2366.62$ \\
   $6_{1/2} $ & $ \Sigma_{c}^{0} $ & $2453.75 \pm 0.14$ & $2445.55$ &$2370.50$
   & $ 2457.83$ & $2366.64$ \\
   & $ \Xi_{c}^{\prime +} $ & $2578.2 \pm 0.5$ & $2579.83$ & $2506.09$
   & $ 2462.42$ & $ 2508.01$ \\
   & $ \Xi_{c}^{\prime 0} $ & $2578.7 \pm 0.5$ & $2580.73$ & $2506.11$ 
   & $ 2549.78$ & $ 2508.04$\\
   & $ \Omega_{c} $ & $2695.2 \pm 1.7$ & $2715.46$ & $2641.28$ 
   & $2641.28$ & $2648.99$ \\
   \hline
   \rule{0in}{2.5ex}
   &  $ \Sigma_{c}^{*++} $ & $2518.41^{+0.21}_{-0.19}$ & $2513.77$
   & $2444.52$ &$2357.18$ &$2440.66$ \\
   & $ \Sigma_{c}^{*+} $ & $2517.5 \pm 2.3 $ & $2512.75$ & $2442.62$ 
   & $2442.62$ & $ 2438.77$ \\
   $6_{3/2} $  & $ \Sigma_{c}^{*0} $ & $2518.48 \pm 0.20 $ & $2513.65 $ 
   & $2442.64$ & $2529.98$ & $2438.79$ \\
   & $ \Xi_{c}^{*+} $ & $2645.10 \pm 0.30 $ & $2647.93$ & $2578.23$
   & $ 2534.57$ & $2580.16$ \\
   & $ \Xi_{c}^{*0} $ & $2646.16 \pm 0.25 $ & $2648.83$ & $2578.26$
   & $2621.92$ & $2580.18$ \\
   & $ \Omega_{c}^{*} $ & $2765.9 \pm 2.0$ & 
   $2784.52$&  $2714.38$&  $2714.38$
  &$2722.09$ \\ [0.1ex]
  \hline \hline
  \end{tabular}
  \end{center}
  \end{table*}
From the fifth column to the last one, we list the results for the
medium-modified values for the masses of the singly charmed
baryons. In the column, their results in symmetric nuclear matter are
listed at the normal nuclear matter density. As expected from the
previous work~\cite{Ghim:2021odo} The masses of the singly charmed
baryons consistently decrease in nuclear matter. We now consider the
mass modification in asymmetric nuclear matter with $\delta =1$
set. Then, as shown in the sixth column, we find a very interesting 
aspect in the change of the $\Xi_c$ masses. In the asymmetric nuclear
matter, the proton and neutron undergo changes in a different manner:
the proton mass starts to decrease as $\delta$ increases, whereas the
neutron mass gets enhanced with larger values of $\delta$. The down 
quarks outnumber the up quarks in asymmetric nuclear matter. If one
puts a down quark in it, the Pauli exclusion principle brings about
the repulsion between the down quarks. Thus, competition between up
and down quarks will govern how the mass of a singly charmed quark is 
modified in asymmetric nuclear matter. It explains why the mass of
$\Xi_c^0$ increases as $\delta$ increases whereas $\Xi_c^+$ behaves
opposedly in asymmetric nuclear matter. A similar propensity can also 
be observed in the baryon sextet, though it is not as
prominent as in the baryon antitriplet. In the last column, we examine
how the masses of the singly charmed baryons experience the medium
modification in strange matter with $\chi=0.15$. As discussed above,
now the number of the strange quarks increases and hence a singly
charmed baryon containing the strange quark may decrease less than the
nonstrange ones. We observe this feature in the last column of
Table~\ref{tab:1}. We will later discuss the density dependences of
the antitriplet and sextet masses quantitatively.

  \begin{table*}[htp]
    \caption{
Masses of the singly bottom baryons in free space and in  
  different baryonic matters at the normal nuclear matter density
  $\lambda=1$.  The experimental data are taken from the
  PDG~\cite{ParticleDataGroup:2022pth}.
  In the fifth column, the results in symmetric nuclear matter
  ($\lambda=1$) are listed, whereas, in the sixth and seventh
  columns, those in asymmetric nuclear matter ($\delta =1$) and
  strange matter ($\chi=0.15$) are respectively given. All the masses
  are given in unit of MeV.}  
    \label{tab:2}
    \begin{center}
    \begin{tabular}{c|ccc|ccc}
    \hline \hline
Multiplet \& spin  & Baryon  & Exp. & Free space &\multicolumn{3}{c}{Baryonic matter 
 at $\lambda=1$.} \\
  &&&&$\delta=0,\,\chi=0$ & $\delta=1,\, \chi=0$ & $\delta=0,\, \chi=0.15$ \\
       \hline 
       \rule{0in}{2.5ex}
      & $ \Lambda_{b} $  & $5619.60 \pm 0.17$ & 
      $5599.30$ & $5595.17$ & $5595.17$ & $5590.95 $ \\
      $\overline{3}_{1/2}$ & $\Xi_{b}^{0}$ & $5791.9 \pm 0.5$ & $ 5800.28$ & $5798.05 $
      & $5736.27 $ & $ 5800.17$ \\
      & $\Xi_{b}^{-}$ & $5797.0 \pm 0.6$ & $5806.02 $ & $5800.00 $
      & $5861.78 $ & $ 5802.11$ \\
     \hline
     \rule{0in}{2.5ex}
     & $ \Sigma_{b}^{+} $ & $5810.56 \pm 0.25$ & $5801.24$ & $5729.83$ 
     & $ 5642.49$  & $5725.98$ \\
     & $ \Sigma_{b}^{0} $ & $- $ & $5802.98$ & $5730.69$ 
     & $ 5730.69$ & $ 5726.84$ \\
     $6_{1/2} $ & $ \Sigma_{b}^{-} $ & $5815.64 \pm 0.18$ & $ 5806.64$
     & $ 5733.48$ & $ 5820.81$ & $ 5729.62$ \\
     & $ \Xi_{b}^{\prime 0} $ & $-$ & $5936.78$ & $5864.93$
     & $ 5821.26$ & $ 5866.85$ \\
     & $ \Xi_{b}^{\prime -} $ & $5935.02 \pm 0.05$ 
     & $ 5940.44$ & $5867.71$ & $ 5911.38$ & $ 5869.64$ \\
     & $ \Omega_{b} $ & $6046.1 \pm 1.7$ & $6074.74$ & $6002.46$ 
     & $ 6002.46$ & $6010.16$ \\
     \hline
     \rule{0in}{2.5ex}
     & $ \Sigma_{b}^{*+} $ & $5830.32 \pm 0.27 $ & $5821.54$ & $5751.34$
     & $5664.00$ & $5747.48$ \\
     & $ \Sigma_{b}^{*0} $ & $-$ & $5823.28$ & $5752.20$ 
     & $ 5752.20 $ & $5748.35$ \\
    $6_{3/2} $  & $ \Sigma_{b}^{*-} $ & $5834.74 \pm 0.30 $ & $ 5826.94$ & $5754.98$ & $5842.32$ & $5751.13$ \\
     & $ \Xi_{b}^{*0} $ & $5952.3 \pm 0.6 $ & $5957.08$ & $5886.43$
     & $5842.77$ & $5888.36$ \\
     & $ \Xi_{b}^{*-} $ & $5955.33 \pm 0.13 $ & $5960.74$ &$5889.22$
     &$5932.88$ & $5891.14$\\
     & $ \Omega_{b}^{*} $ & $-$ &$6095.04$
     & $6023.96$ &  $6023.96$ &$6031.67$ \\ [0.1ex]
    \hline \hline
    \end{tabular}
    \end{center}
    \end{table*}
For completeness, we list the results for the mass modification of the
singly bottom baryons in Table~\ref{tab:2}. Except for the spin-spin
interaction that is proportional to $1/m_Q$, we respect in the current
work the heavy-quark flavor symmetry. Thus, the changes of the masses
of the singly bottom baryons are in conformity with those of the
charmed baryons. 
    
\begin{figure}[htpb!]
  \begin{center}
  \includegraphics[scale=0.17]{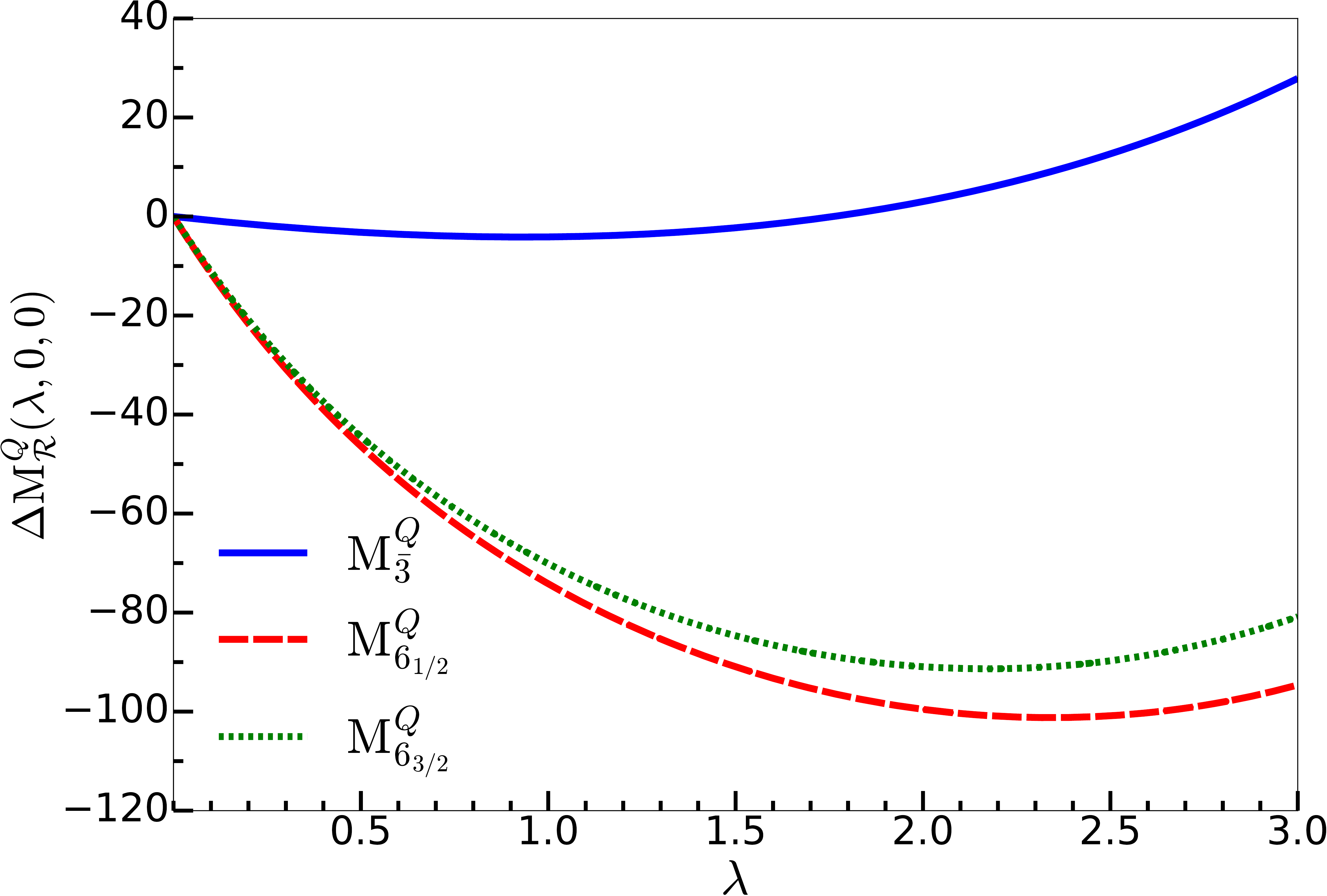} 
  \end{center}
  \caption{Shifts of the center masses for the singly heavy
    baryons. The solid curve draws the mass shift of the baryon
    antitriplet. The dashed and dotted ones depict
    respectively the mass shifts for the baryon sextet with spin 1/2
    and spin 3/2. The results are given in unit of MeV.}  
  \label{fig:1}
\end{figure}
\begin{figure*}[htpb!]
  \includegraphics[scale=0.17]{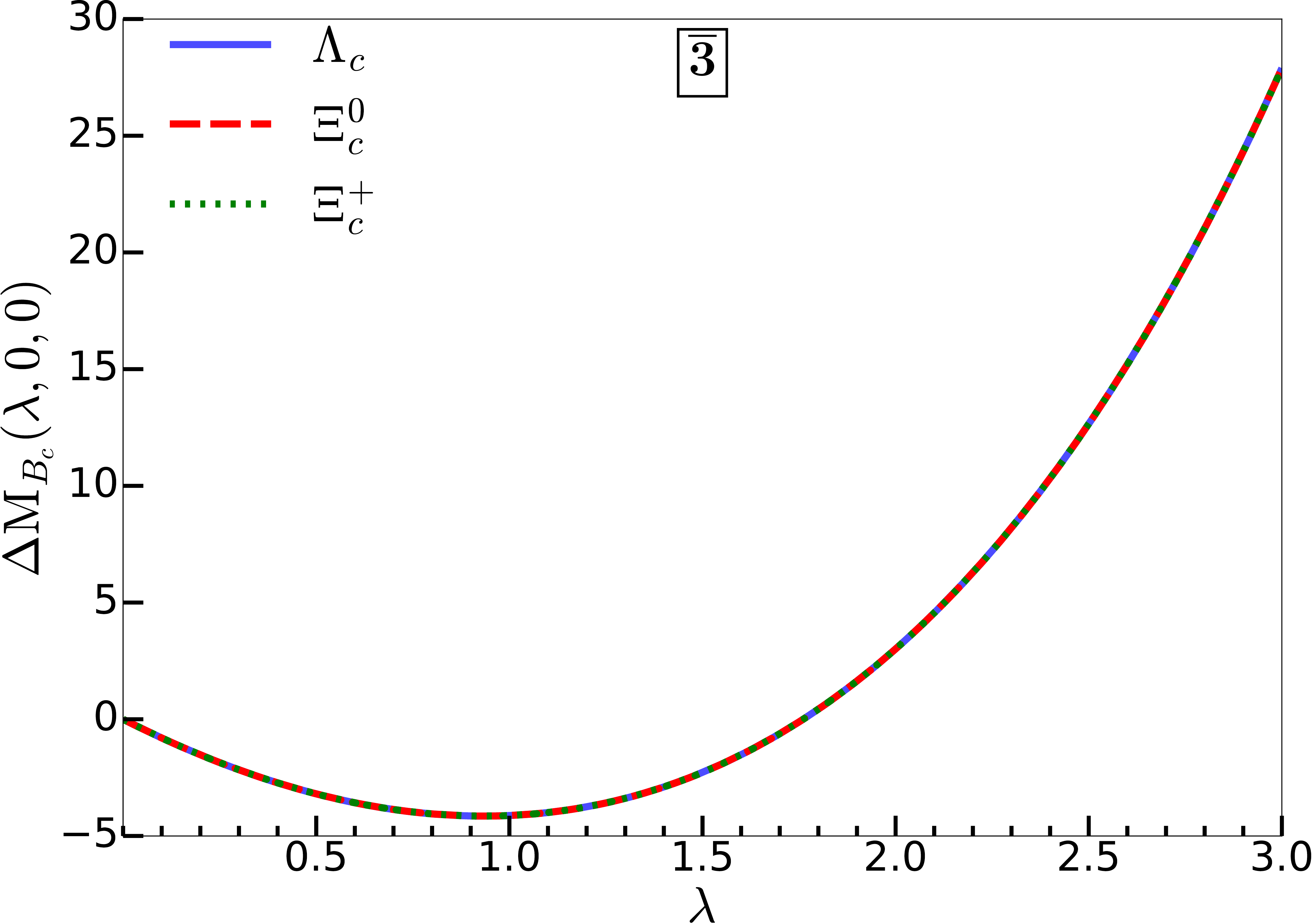}\;\;\;
  \includegraphics[scale=0.17]{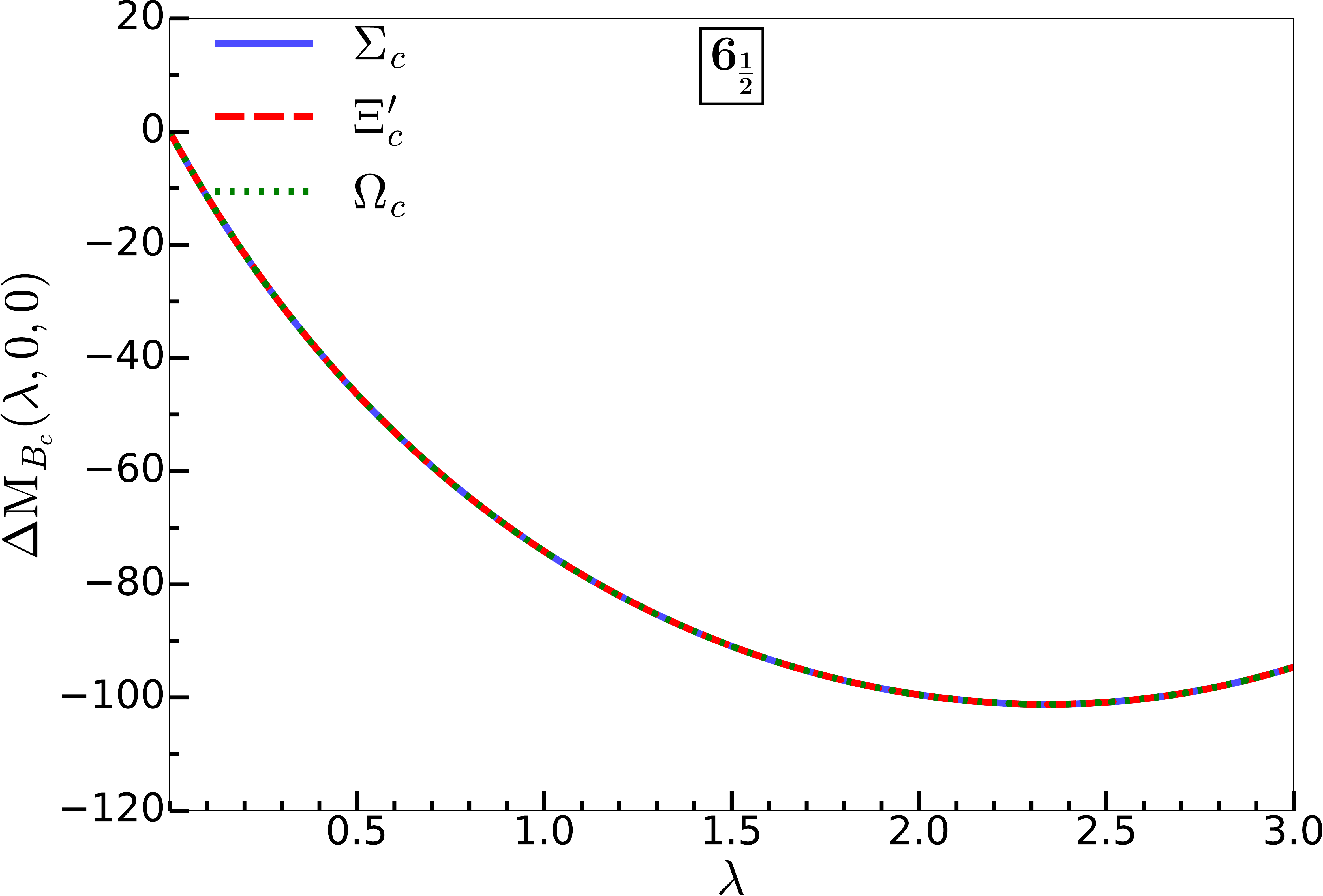}
  \includegraphics[scale=0.17]{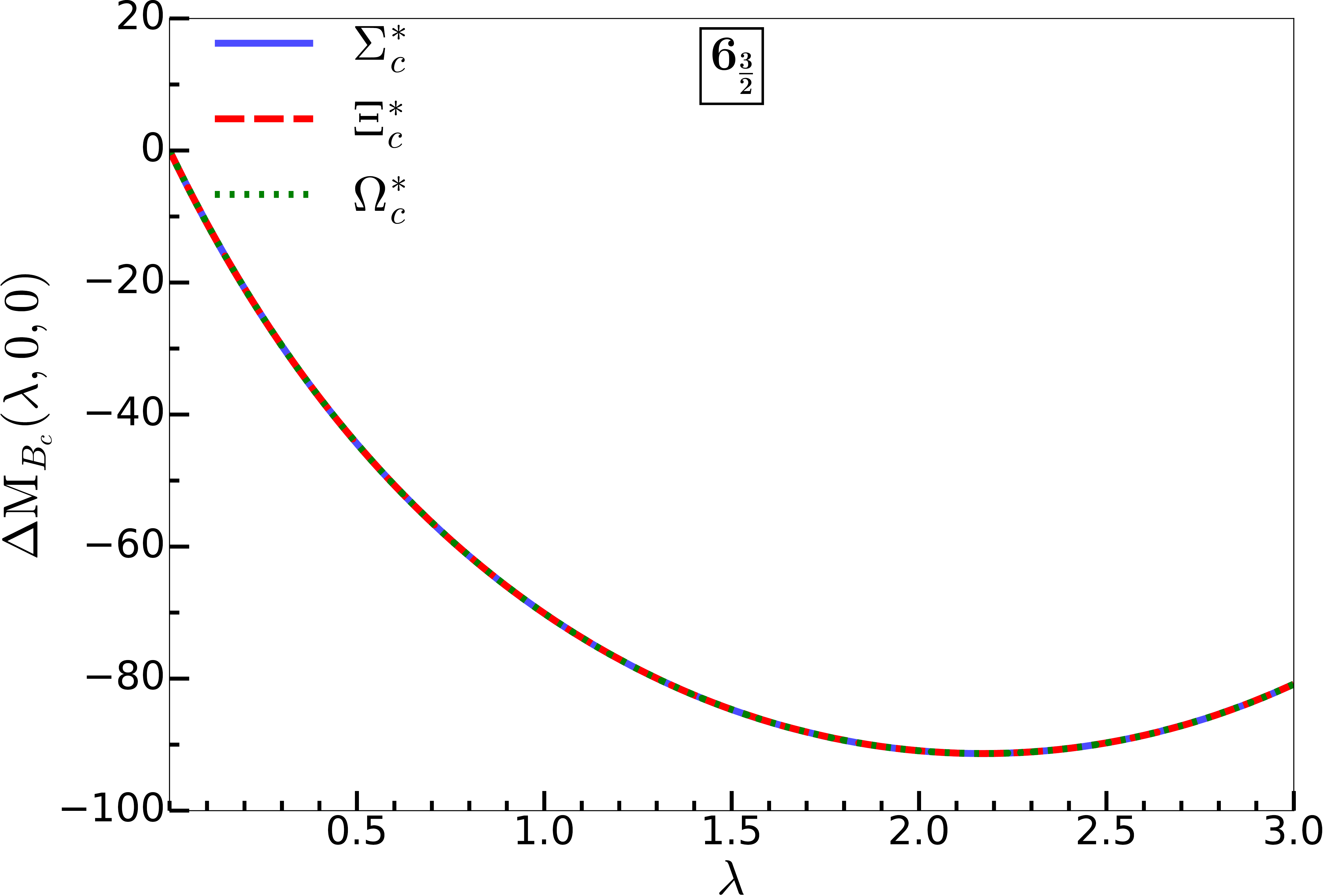}
  \caption{Mass shifts of singly charmed baryons in symmetric nuclear
    matter ($\delta=1$, $\chi=0$). In the upper left panel, the
    $\lambda$ dependences of the baryon antitriplet are drawn. In the
right upper panel, those of the baryon sextet with spin 1/2 are
    depicted, whereas in the lower panel, those of the baryon sextet
    with spin 3/2 are shown. The results are given in unit of
    MeV.} 
  \label{fig:2}
\end{figure*}
Figure~\ref{fig:1} draws the mass shifts of the center masses, i.e.,
$\Delta M_c^{\mathcal{R}}= M_c^{\mathcal{R}*}-M_c^{\mathcal{R}}$,
where the superscript $\mathcal{R}$ denotes the corresponding
representation. The expressions for $M_c^{\mathcal{R}}$ are given in
Eqs.~\eqref{eq:26} and~\eqref{eq:30}, as functions of $\lambda$.
Note that the center 
$\Delta M_c^{\overline{\bm{3}}}$ decreases as $\lambda$ increases till
$\lambda \approx 1.2$, and then gets enhanced. On the other hand, 
$\Delta M_c^{\bm{6}_{1/2}}$ ($\Delta M_c^{\bm{6}_{3/2}}$) is
diminished rapidly till $\lambda$ reaches around 2.2 (2.5) and then
starts to increase. It implies that when the nucleons inside nuclear
matter get more closely packed the repulsion overcomes the attractive
interaction in the presence of the singly charmed baryons. The
difference between the density dependences of the antitriplet and
sextet can be understood as follows: the density dependences of
$\overline{I}_1$ and $\overline{I}_2$ are different each other. While
$\overline{I}_1$ increases as $\lambda$ increases, $\overline{I}_2$ is
lessened with the $\lambda$ grown. $M_{\mathrm{cl}}$ decreases
linearly as the nuclear density increases. When $\lambda$ reaches
around 1.2, the second term $1/2\overline{I}_2$ overtakes
$M_{\mathrm{cl}}$, so that $M_{\overline{\bm{3}}}$ starts to
increase. However, the second term  
for $M_{\bm{6}_{1/2}}$ and $M_{\bm{6}_{3/2}}$ in ~\eqref{eq:26} is
suppressed as $\lambda$ increases. Thus, $M_{\bm{6}_{1/2}}$ and
$M_{\bm{6}_{3/2}}$ follow the behavior of $M_{\mathrm{cl}}$. When
$\lambda$ further increases, the term with $\varkappa^*$ comes into
play. In Fig.~\ref{fig:2}, we draw the mass shifts of the charmed
baryon antitriplet and sextet, $\Delta M_{B_c}$, in symmetric nuclear
matter. The $\lambda$ dependences of $\Delta M_{B_c}$ follow those of
the center masses shown in Fig.~\ref{fig:1}. This is natural, because
the effects of the flavor SU(3) symmetry breaking, which causes the
mass splitting in the representations, are changed only in strange
matter. This is the reason why the mass shift in each representation
is degenerate. 

\begin{figure*}[htpb!]
  \includegraphics[scale=0.17]{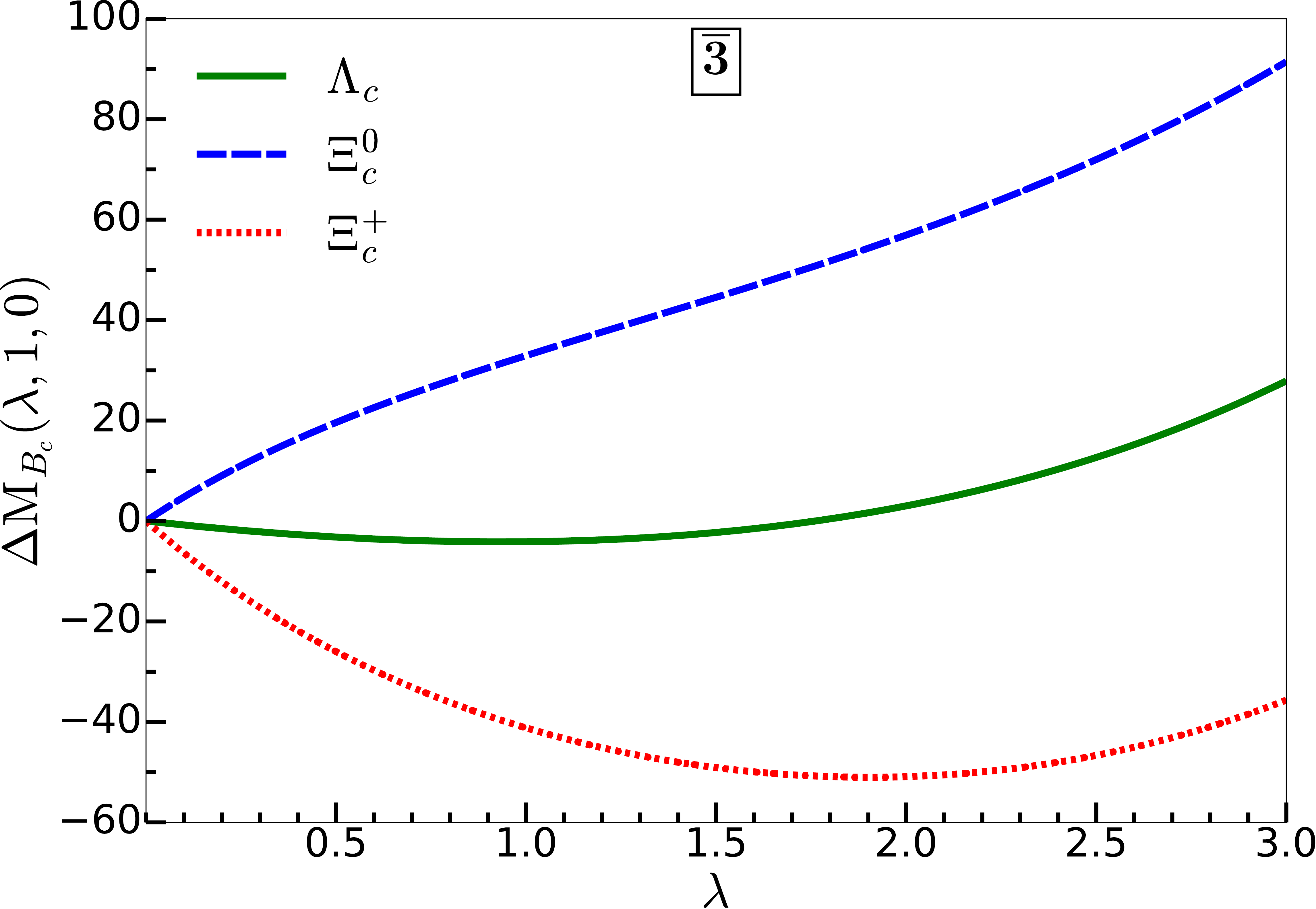}\;\;\;
  \includegraphics[scale=0.17]{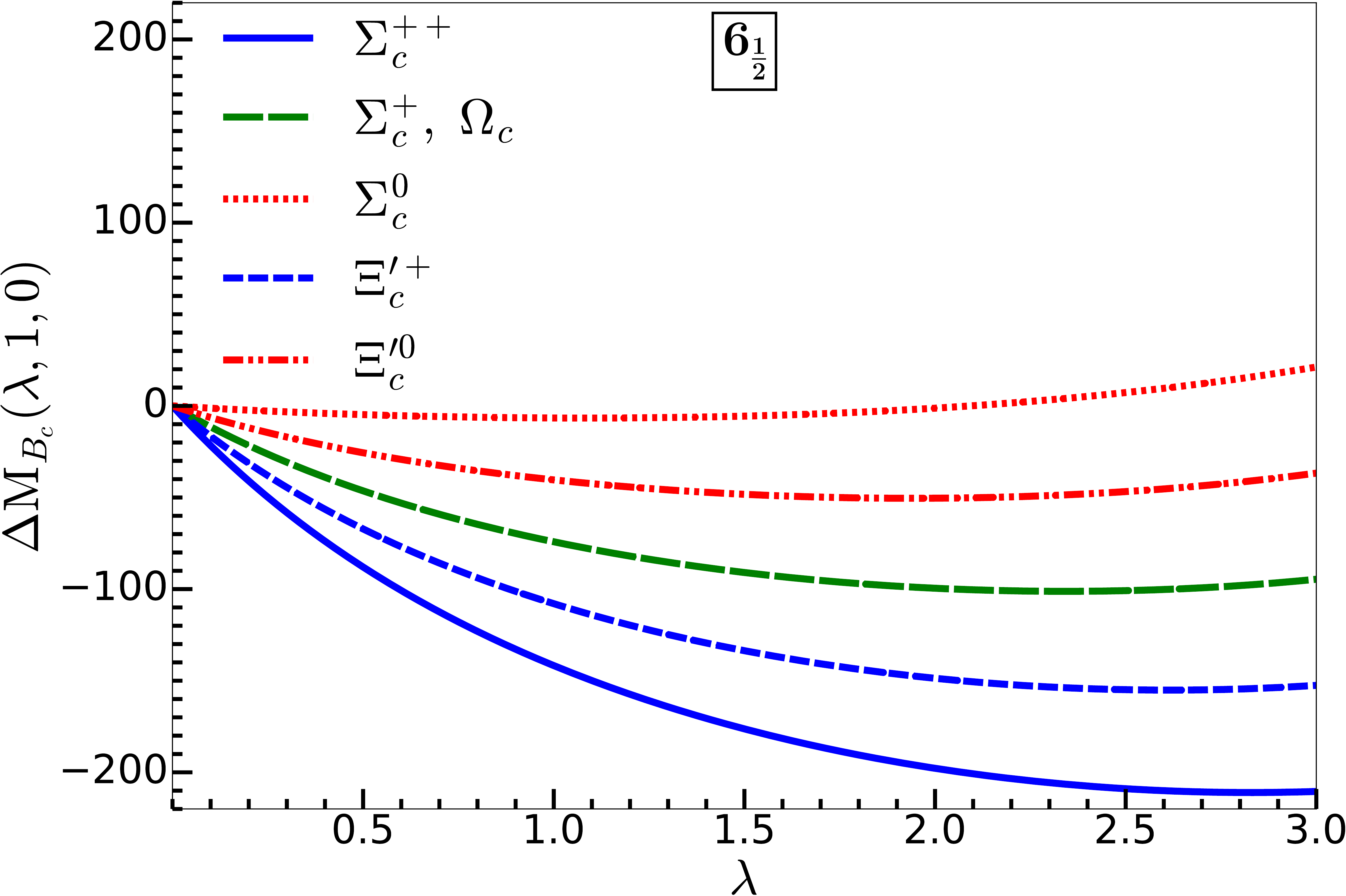}
  \includegraphics[scale=0.17]{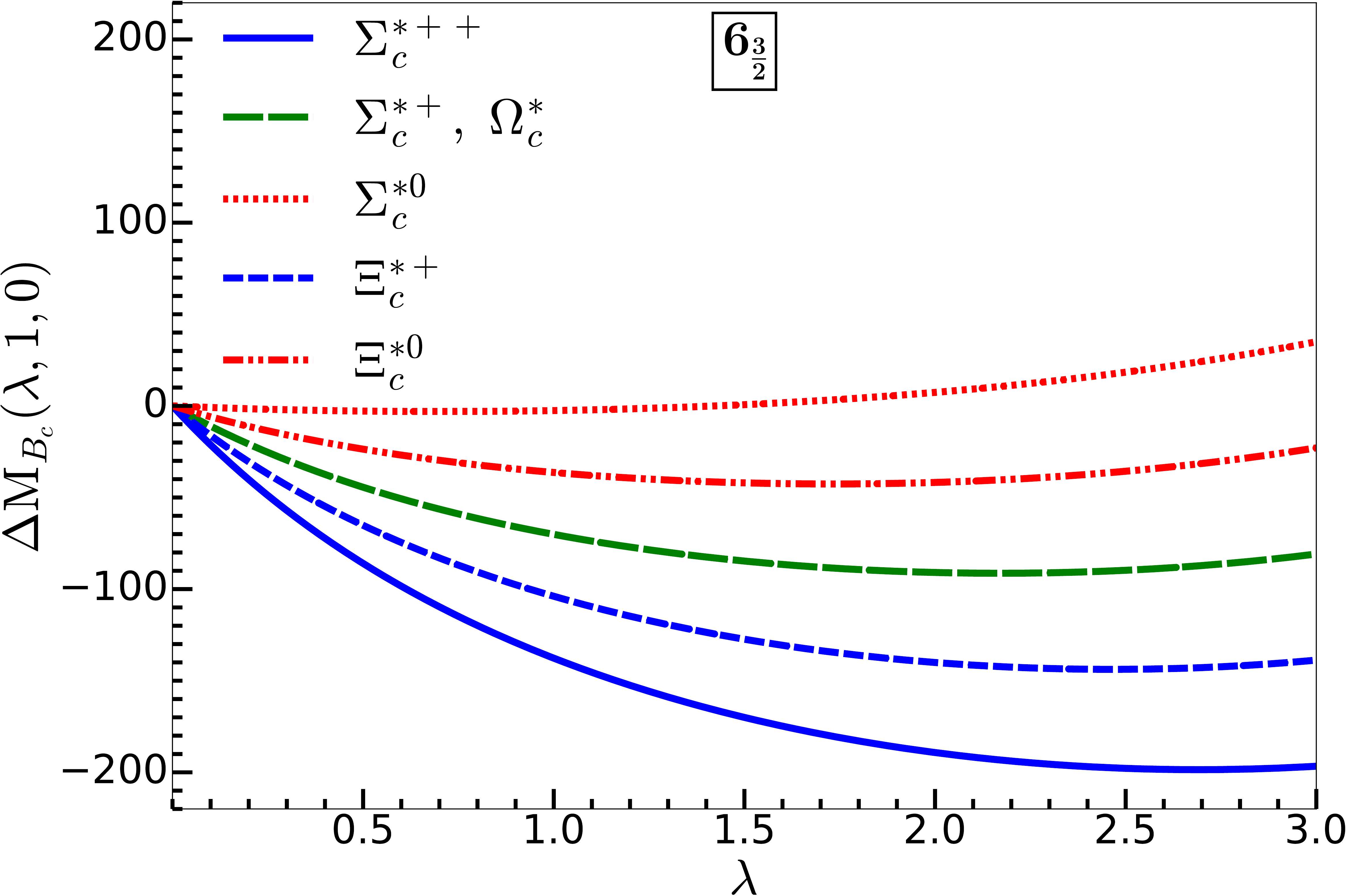}
  \caption{Mass shifts of singly charmed baryons in asymmetric nuclear
    matter ($\delta=1$, $\chi=0$). In the upper left panel, the
    $\lambda$ dependences of the baryon antitriplet, i.e.,
    $\Lambda_c$, $\Xi_c^0$, and $\Xi_c^+$ are drawn in solid curve,
    dashed one, and dotted one, respectively. In the right 
    upper panel, those of the baryon sextet with spin 1/2 are
    depicted, whereas in the lower panel, those of the baryon sextet
    with spin 3/2 are shown. The results are given in unit of MeV.} 
  \label{fig:3}
\end{figure*}
In Fig.~\ref{fig:3}, we depict the mass shifts of the singly charmed
baryons in asymmetric neutron matter with $\delta =1$ and $\chi=0$. 
The neutral and positively-charged baryons generally show rather
different behaviors as $\lambda$ increases. The charmed baryons in the 
antitriplet exhibit the difference prominently. While $\Xi_c^0$
increases rather rapidly as $\lambda$ increases, $\Xi_c^+$ decreases
until $\lambda$ reaches around $\lambda=2.0$ in asymmetric nuclear
matter. It indicates that the effects of isospin symmetry breaking
stand out in neutron matter ($\delta=1$). This has profound physical
implications. The density-dependent function $f_0(\lambda,1,0)$ in
Eq.~\eqref{eq:f0} increases as $\delta$ grows. It contributes to the
$\overline{\alpha}$, $\beta$, and $\gamma$ in 
Eq.~\eqref{eq:alphabetagamma}, so that $d_3$ and $d_6$ in
Eq.~\eqref{eq:brpar} become $\delta$-dependent. The terms containing
$d_3$ $d_6$ in mass formulae in
Eqs.~\eqref{eq:M3bar}--\eqref{eq:omegastr} are proportional to the
third component of the isospin operator, $T_3$, which brings about the
isospin symmetry breaking. Thus, the differences between the neutral
and positively charged baryons demonstrated in Fig.~\ref{fig:3} arise
from these terms. As explained above, the underlying physics in these
differences comes from the Pauli exclusion principle.  

\begin{figure*}[htpb!]
    \includegraphics[scale=0.17]{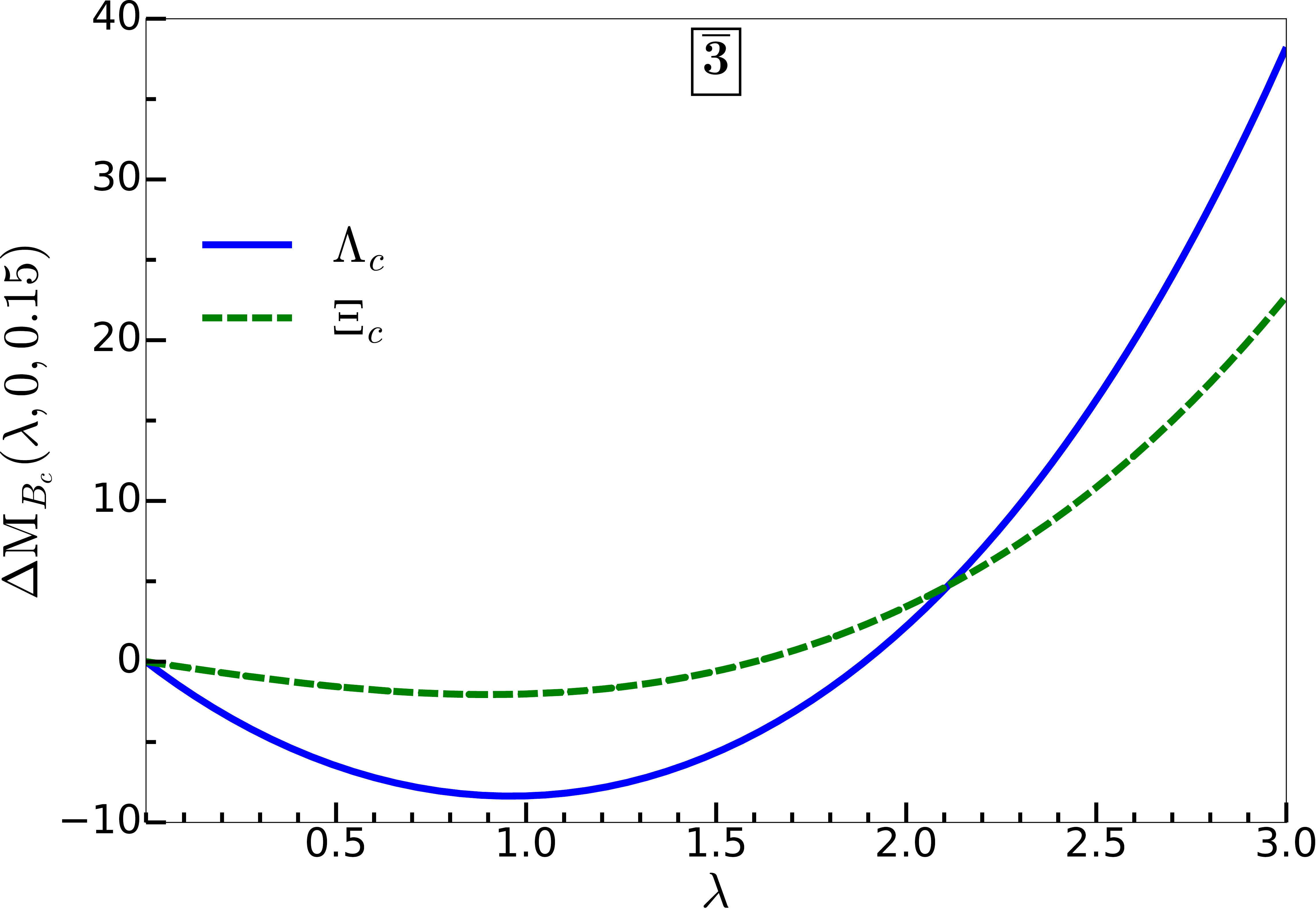}\;\;\;
    \includegraphics[scale=0.17]{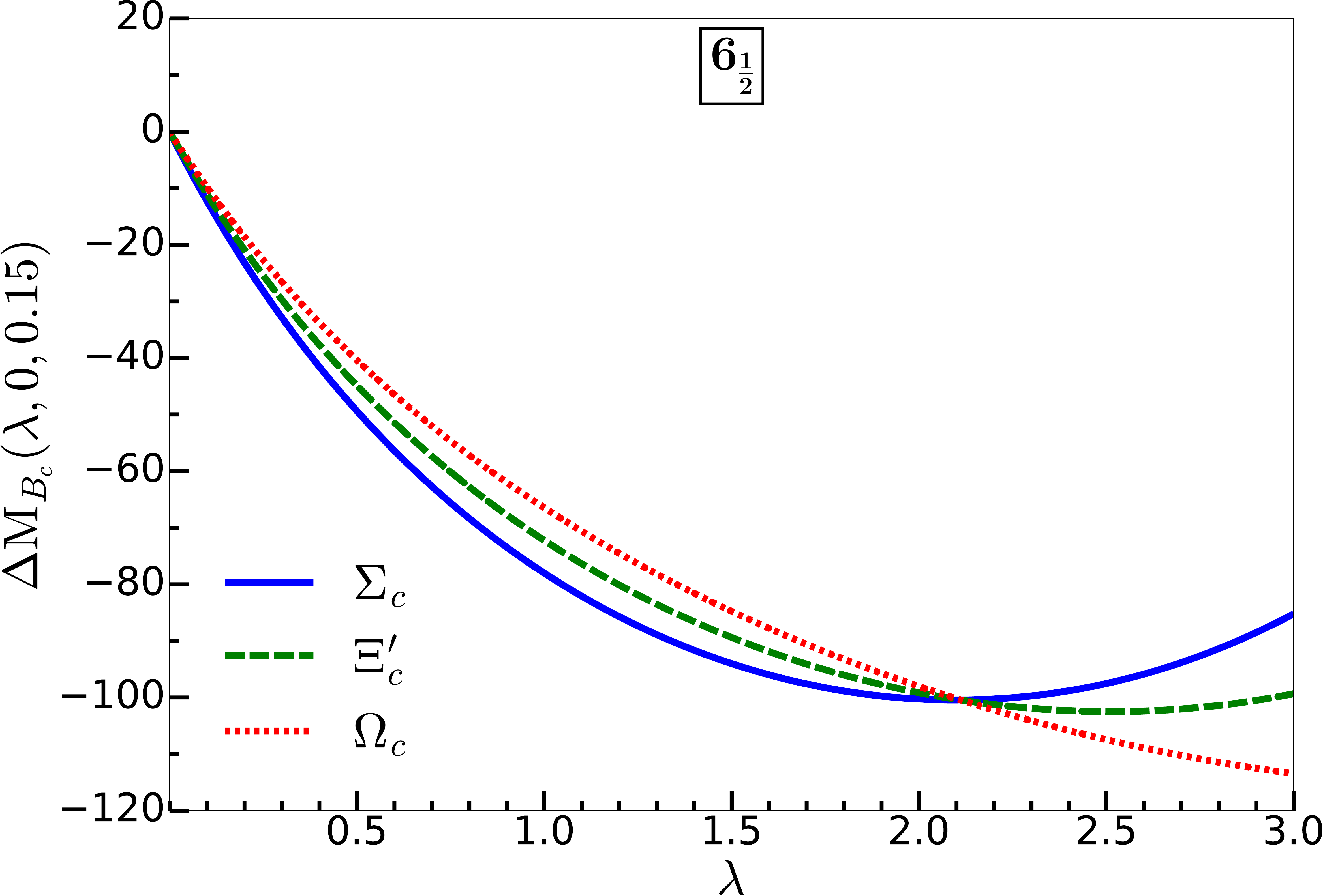}
    \includegraphics[scale=0.17]{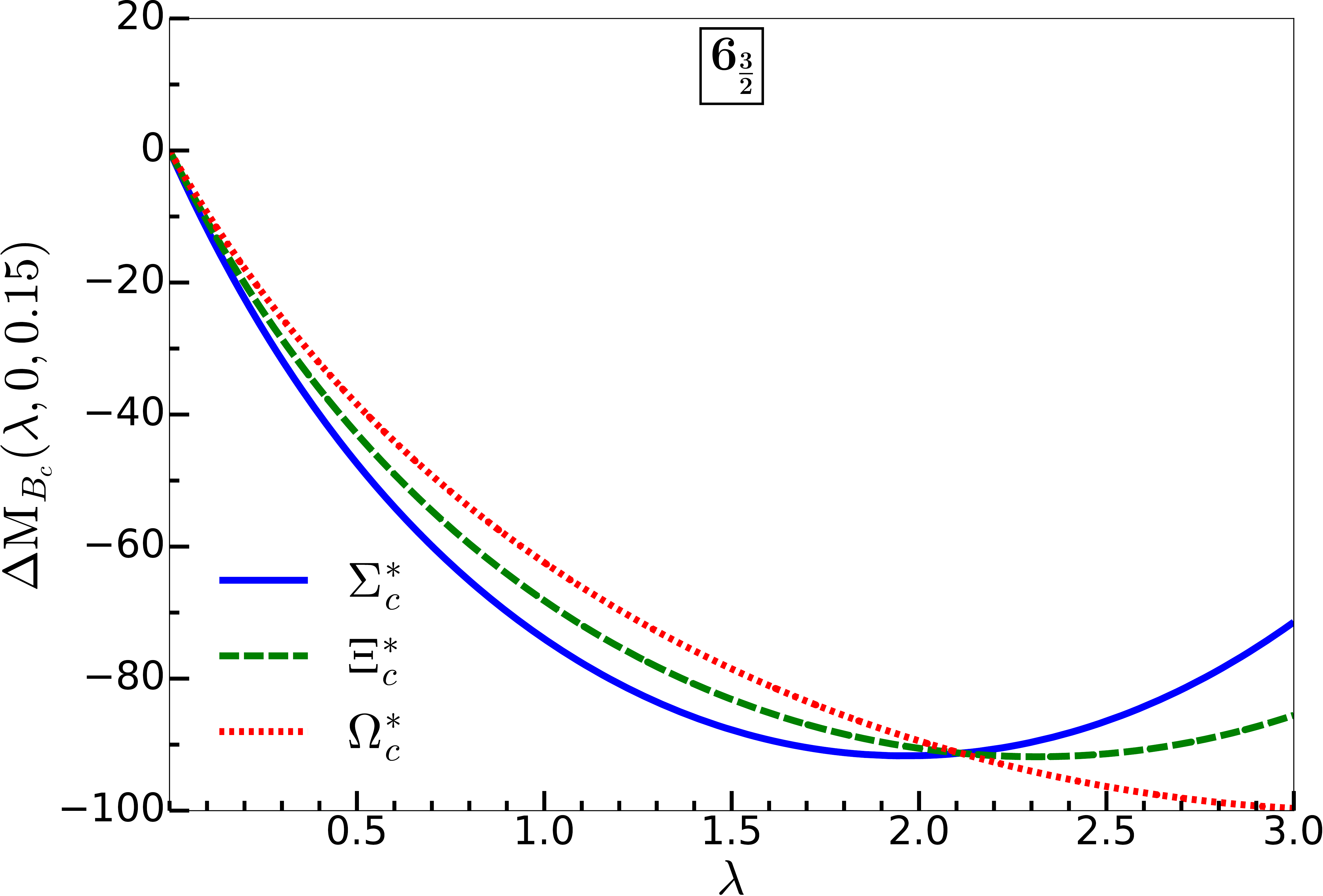}
    \caption{Mass shifts of singly charmed baryons in strange matter
    ($\delta=0$, $\chi=0.15$). In the upper left panel, the $\lambda$
    dependences of the baryon antitriplet, i.e., $\Lambda_c$ and
    $\Xi_c$ are drawn in solid curve and dashed one,
    respectively. In the right upper panel, those of the baryon sextet
    with spin 1/2 are depicted, whereas in the lower panel, those
    of the baryon sextet with spin 3/2 are shown. 
    The results are given in unit of MeV.}
    \label{fig:4}
\end{figure*}    
Figure~\ref{fig:4} illustrates how the masses of the singly charmed
baryons are shifted as $\lambda$ increases. Interestingly, 
the mass shifts of the singly charmed baryon show general tendency:
They first start to decrease as $\lambda$ increases, and then
increases when $\lambda$ gets to some specific values. However, those
of $\Omega_c$ and $\Omega_c^*$ monotonically fall off as $\lambda$
increases. Inspecting Eqs.~\eqref{eq:M3bar}--\eqref{eq:omegastr}, we
find that the terms with $D_3$ in the antitriplet and $D_6$ in the
sextet cause respectively the mass splittings in the corresponding
representations. We also observe that the $\lambda$ dependences of the
baryon sextet with spin 1/2 are almost the same as those with spin
3/2. Note that the sextet baryons with spin 1/2 and 3/2 are degenerate
before we introduce the hyperfine interaction in
Eq.~\eqref{eq:30}. Though the parameter $\varkappa^*$ in
Eq.~\eqref{eq:den_chi} is also density-dependent, its effect is
marginal. The prefactor in the $D_6$ term of the $\Omega_c$
($\Omega_c^*$) baryons is $-4/3$, whereas the those of $\Sigma_c$
($\Sigma_c^*$) and $\Xi_c'$ ($\Xi_c^*$) are respectively $+2/3$ and
$-1/3$. This leads to the different $\lambda$ dependences of the
sextet baryons as shown in Fig.~\ref{fig:4}.

\section{Summary and outlook\label{sec:5}}
In the present work, we aimed at investigating the mass shifts of the
singly heavy baryons within a pion mean-field approach ($\chi$QSM) in
various nuclear matters. In the limit of the infinite heavy-quark
mass, the dynamics in a singly heavy baryon is governed by the light
quarks whereas the heavy remains as the mere static color source with
the heavy quark spin-flavor symmetry satisfied. The light quarks,
which yields the right hypercharge $Y'=2/3$, select the proper
representations of the singly heavy baryons. This allows one to
describe the light and singly heavy baryons on an 
equal footing. Since all the density-dependent variables had been 
determined in describing the bulk properties of nuclear matter and the
mass shifts of the baryon octet and decuplet, we were able to evaluate
those of the baryon antitriplet and sextet without fitting the
parameters. Then, we first computed the medium-modified masses of the
singly charmed baryons in symmetric nuclear matter. The center masses
of the baryon antitriplet and sextet govern the density dependences of 
the singly charmed baryon masses. In the case of asymmetric nuclear
matter, the neutral and positively-charged baryons reveal different
density dependences: The neutral baryons tend to increase as the
nuclear density increases, whereas the positively charged ones
decrease as the nuclear density grows. We explained the reason and
discussed its physical implications. As a result, the effects of
isospin symmetry breaking are more strengthened as the density
increases in asymmetric nuclear matter. We also presented the mass
shifts of the singly charmed baryons in strange matter.

\section*{Acknowledgments}
The present work was supported by Basic Science Research Program
through the National Research Foundation of Korea funded by the
Ministry of Education, Science and Technology
(Grant-No. 2021R1A2C209336 and 2018R1A5A1025563 (H.-Ch. K.),
and 2020R1F1A1067876 (U. Y.)). 

\appendix

\section{Expressions for the masses of the singly
  heavy baryons} 
\label{app:a}
The masses of the antitriplet baryon are expressed as
\begin{align}
  M_{\Lambda_{\mathrm{Q}}}=& M_{\mathrm{cl}}+E^{\mathrm{rot}}_{(0,1)}
  +m_{\mathrm{Q}}+\frac{2}{3}D_{3}+\frac{1}{4}c_{8},\cr
  M_{\Xi_{\mathrm{Q}}}=&M_{\mathrm{cl}}+E^{\mathrm{rot}}_{(0,1)}
  +m_{\mathrm{Q}}-\frac{1}{3}D_{3}+d_{3}T_{3}\cr
  &+\frac{3}{4}\left(T_{3}+\frac{1}{6}\right)c_{8} -
    \hat{Q}_{q}\alpha_{LQ} T_{3}, 
  \label{eq:M3bar}
\end{align}
where the $E^{\mathrm{rot}}_{0,1}$ can be obtained from
Eq~\eqref{eq:Erot}. The masses of spin-1/2 sextet baryon are given by
following espression: The masses of 
\begin{align}
  M_{\Sigma_{\mathrm{Q}}}=&M_{\mathrm{cl}}+E^{\mathrm{rot}}_{(2,0)}
  +m_{\mathrm{q}}-\frac{2}{3}\frac{\varkappa}{m_{\mathrm{Q}}}\cr
  &+\frac{2}{3}D_{6}+d_{6}T_{3}+\frac{3}{10}\left( T_{3}+\frac{1}{3}
    \right)c_{8}\cr 
  &+\frac{1}{9}\left( T_{3}^{2}+\frac{1}{5}T_{3} -\frac{3}{5}\right)c_{27}\cr
  &+\hat{Q}_{q}\alpha_{LQ}T_{3},
\label{eq:sigma}
    \end{align}
    \begin{align}
    M_{\Xi'_{\mathrm{Q}}}=&M_{\mathrm{cl}}+E^{\mathrm{rot}}_{(2,0)}
  +m_{\mathrm{Q}}-\frac{2}{3}\frac{\varkappa }{m_{\mathrm{Q}}}\cr
  &-\frac{1}{3}D_{6}+d_{6}T_{3}+\frac{3}{10}\left( T_{3} -\frac{1}{6} \right)c_{8},\cr
  &-\frac{2}{45}\left(
    T^{2}_{3}+2T_{3}+\frac{1}{4}\right)c_{27}+\hat{Q}_{q}\alpha_{\mathrm{LQ}}T_{3},
\label{eq:xi}
\end{align}
\begin{align}
    M_{\Omega_{\mathrm{Q}}}=&M_{\mathrm{cl}}+E_{(2,0)}^{\mathrm{rot}}
      +m_{\mathrm{Q}}-\frac{2}{3}\frac{\varkappa }{m_{\mathrm{Q}}}
    -\frac{4}{3}D_{6}\cr  
  &+\frac{1}{5}c_{8}-\frac{1}{45}c_{27}.
\label{eq:omega}
\end{align}
where the $E^{\mathrm{rot}}_{2,0}$ can be obtained from
Eq~\eqref{eq:Erot}. The masses of spin-3/2 baryon sextet mass can be
written as following expression : 
\begin{align}
  M_{\Sigma^{*}_{\mathrm{Q}}}=&M_{\mathrm{cl}}+E_{(2,0)}^{\mathrm{rot}}
  +m_{\mathrm{Q}}-\frac{1}{3}\frac{\varkappa }{m_{\mathrm{Q}}}\cr
  &+\frac{2}{3}D_{6}+d_{6}T_{3} +\frac{3}{10}\left( T_{3} +
    \frac{1}{3} \right)c_{8}\cr
  &+\frac{1}{9}\left( T_{3}^{2}+\frac{1}{5}T_{3}
    -\frac{3}{5} \right)c_{27}\cr
  &+ \hat{Q}_{q}\alpha_{LQ}T_{3},
\label{eq:sigmastr}
    \end{align}
    \begin{align}
    M_{\Xi_{\mathrm{Q}}^{*}}=& M_{\mathrm{cl}}
       + E_{(2,0)}^{\mathrm{rot}}+m_{\mathrm{Q}} -
    \frac{1}{3}\frac{\varkappa }{m_{\mathrm{Q}}}\cr
  &-\frac{1}{3}D_{6}+d_{6}T_{3}+\frac{3}{10}
    \left( T_{3}-\frac{1}{6} \right)c_{8}\cr
  &-\frac{2}{45}\left( T_{3}^{2}+2T_{3}+\frac{1}{4} \right)c_{27}\cr
      &+\hat{Q}_{q}\alpha_{LQ}T_{3},
\label{eq:xistr}
\end{align}
\begin{align}
    M_{\Omega^{*}_{\mathrm{Q}}}=& M_{\mathrm{cl}}
            +E_{(2,0)}^{\mathrm{rot}}+m_{\mathrm{Q}}
          -\frac{1}{3}\frac{\varkappa }{m_{\mathrm{Q}}}\cr
  &-\frac{4}{3}D_{6}+\frac{1}{5}c_{8}-\frac{1}{45}c_{27}.
\label{eq:omegastr}
\end{align}
Here $d_{3,6}$ and $D_{3,6}$ are defined as
\begin{align}
  D_{3}=&\left( m_{s}-\hat{m} \right)\left(
          \frac{3}{8}\overline{\alpha}+\beta \right),\cr 
  D_{6}=&\left( m_{s}-\hat{m} \right)\left(
          \frac{3}{20}\overline{\alpha}+\beta-\frac{3}{10}\gamma
          \right),\cr 
  d_{3}=&\left( m_{d}-m_{u} \right)\left( \frac{3}{8}\overline{\alpha}
          +\beta\right),\cr 
  d_{6}=&\left( m_{d}-m_{u} \right)\left(
          \frac{3}{20}\overline{\alpha}
          +\beta-\frac{3}{10}\gamma\right). 
\label{eq:brpar}
\end{align}

\bibliography{HB_NM}

\end{document}